\documentclass{svproc}
\usepackage{graphicx,subfigure}
\usepackage{amsfonts}
\usepackage{url}

\begin{document}
\mainmatter              
\title{MG-VAE: Deep Chinese Folk Songs Generation with Specific Regional Style}
\titlerunning{MG-VAE}  
%
\author{Jing Luo\inst{1} \and Xinyu Yang\inst{1}
Shulei Ji\inst{1} \and Juan Li\inst{2}}
\authorrunning{Jing Luo et al.} 
%
\tocauthor{Jing Luo, Xinyu Yang, Shulei Ji, and Juan Li}
\institute{School of Computer Science and Technology, Xi'an Jiaotong University, Xi'an, China
\and
Center of Music Education, Xi'an Jiaotong University, Xi'an, China}

\maketitle              

\begin{abstract}
Regional style in Chinese folk songs is a rich treasure that can be used for ethnic music creation and folk culture research. In this paper, we propose MG-VAE, a music generative model based on VAE (Variational Auto-Encoder) that is capable of capturing specific music style and generating novel tunes for Chinese folk songs (\emph{Min Ge}) in a manipulatable way. Specifically, we disentangle the latent space of VAE into four parts in an adversarial training way to control the information of pitch and rhythm sequence, as well as of music style and content. In detail, two classifiers are used to separate style and content latent space, and temporal supervision is utilized to disentangle the pitch and rhythm sequence. The experimental results show that the disentanglement is successful and our model is able to create novel folk songs with controllable regional styles. To our best knowledge, this is the first study on applying deep generative model and adversarial training for Chinese music generation.
\end{abstract}
\section{Introduction}
Creating realistic music pieces automatically has always been regarded as one of frontier subjects in the field of computational creativity. With recent advances in deep learning, deep generative model and its variants have been widely used in automatic music generation \cite{HerremansCC17}\cite{deepgenerationsurvey}. However, most of deep composition methods focus on Western music rather than Chinese music. How to employ deep learning to model the structure and style of Chinese music is a challenging but novel problem.

Chinese folk songs, an important part of traditional Chinese music, are improvised by local people and passed on from one generation to the next orally. Folk tunes from the same region exhibit similar style while tunes from different areas present different regional styles \cite{Miao1985}\cite{LiLDZY19}. For example, the songs named \emph{Mo Li Hua} have different versions in many areas of China and show various music styles, though they share the same name and similar lyrics \footnote{The Chorus of \emph{Mo Li Hua Diao} from various regions in China by Central National Orchestra: \url{http://ncpa-classic.cntv.cn/2017/05/11/VIDEEMEg82W5MuXUMM1jpEuL170511.shtml}.} . The regional characteristics of Chinese folk songs are not well explored and should be utilized to guide automatic composition for Chinese folk tunes. Furthermore, folk song composition based on regional style provides abundant potential materials for Chinese national music creation, and promotes the spread and development of Chinese national music and even Chinese culture in the world.

There are lots of studies on music style composition of Western Music\cite{Dai2018}.  However, few studies employ deep generative model for Chinese music composition. There is a clear difference between Chinese and Western music. Unlike Western Music, which focuses on the vertical structure of music, Chinese music focuses on the horizontal structure, i.e., the development of melody, and the regional style of Chinese folk songs is mainly reflected in its rhythm and pitch interval patterns \cite{Guan2014}. 

In this paper, we propose a deep music generation model named MG-VAE to capture regional style of Chinese folk songs (\emph{Min Ge}) and create novel tunes with controlled regional style. Firstly, a MIDI dataset with more than 2000 Chinese folk songs covering six regions is collected. After that, we encode the input music representations to the latent space and decode the latent space to reconstruct music notes. In detail, the latent space is divided into two parts to present the pitch features and rhythm features, namely, \emph{pitch variable} and \emph{rhythm variable}. Then we further divide the pitch latent space into \emph{style variable} part and \emph{content variable} part to present style feature and style-less feature in pitch variable, the same operation is launched in rhythm variable.  In order to capture the regional style of Chinese folk songs precisely and generate regional style songs in controllable way, we propose a method based on adversarial training for disentanglement of the four latent variables, where temporal supervision is employed in the separation of pitch and rhythm variable, and label supervision is used for the disentanglement the style and content variable. The experimental results and visualization of latent spaces show that our model is effective to disentangle latent variables and is able to generate folk songs with specific regional style. 

The rest of the paper is structured as follows: after introducing related work on deep music generation in Section 2, we present our music representations and model in Section 3. Section 4 describes the experimental results and analysis of our methods. Conclusions and future work are presented in Section 5.

\section{Related Work}
RNN (Recurrent Neural Network) is one of the most earliest models introduced into the domain of deep music generation. Researchers employ RNNs to model the music structure and generate different formats of music, including monophonic folk melodies \cite{SturmSBK16}, rhythm composition \cite{MakrisKKK19}, expressive music performance \cite{Sageev19}, multi-part music harmonization \cite{YanLVD18}. Other recent studies have started to combine convolutional structure and explore using VAE, GAN (Generative Adversarial Network) and Transformer for music generation. MidiNet \cite{YangCY17} and MuseGAN \cite{DongHYY18} combine CNN (Convolutional Neural Network) and GAN architecture to generate music with multiple MIDI tracks. MusicVAE \cite{RobertsERHE18} introduces a hierarchical decoder into general VAE model to generate music note sequences with long-term structure. Due to the impressive results of Transformer in neural translation, Huang et al. modify this sequence model’s relative attention mechanism and generate minutes of music clips with high long-range structural coherence \cite{MusicTransformer}. In addition to the study of music structure, researchers also employ deep generative models to model music styles, such as producing jazz melodies through two LSTM networks \cite{JohnsonKW17}, harmonizing a user-made melody in Bach’s style \cite{Coconet2019}. Most of them are trained on the specific style dataset. The music generated from these models can only mimic the single style embodied in the training data. 

Moreover, little attention has been paid to Chinese music generation with deep learning techniques, especially for modeling the music style of Chinese music, though some researchers utilize Seq2Seq model to create multi-track Chinese popular songs from scratch \cite{ZhuLYQLZZWXC18} or generate melody of Chinese popular songs with given lyrics \cite{Bao2018}. The existing generation algorithms for Chinese traditional songs are mostly based on non-deep models such as Markov models \cite{HuangLNC16}, genetic algorithms \cite{ZhengWLSGGW17}. These studies cannot break up the bottleneck in melody creation and style imitation.  

Some latest work in the domain of music style transfer begins to generate music with mixed style or recombine music content and style.  For example, Mao et al. propose an end to end generative modal to produce music with mixture of different classical composer styles \cite{MaoSC18}. Lu et al. study the deep style transfer between Bach chorales and Jazz \cite{LuS18a}. Nakamura et al. complete melody style conversion among different music genres \cite{NakamuraSNY19}. The above studies are based on the music data from different genres or composing periods. However, the regional style generation of Chinese folk songs studied here is modeling style within the same genre, which is more challenging.

\section{Approach}

\subsection{Music Representation}
\begin{figure}[htb]
	\centering
	\includegraphics[width=3.6in]{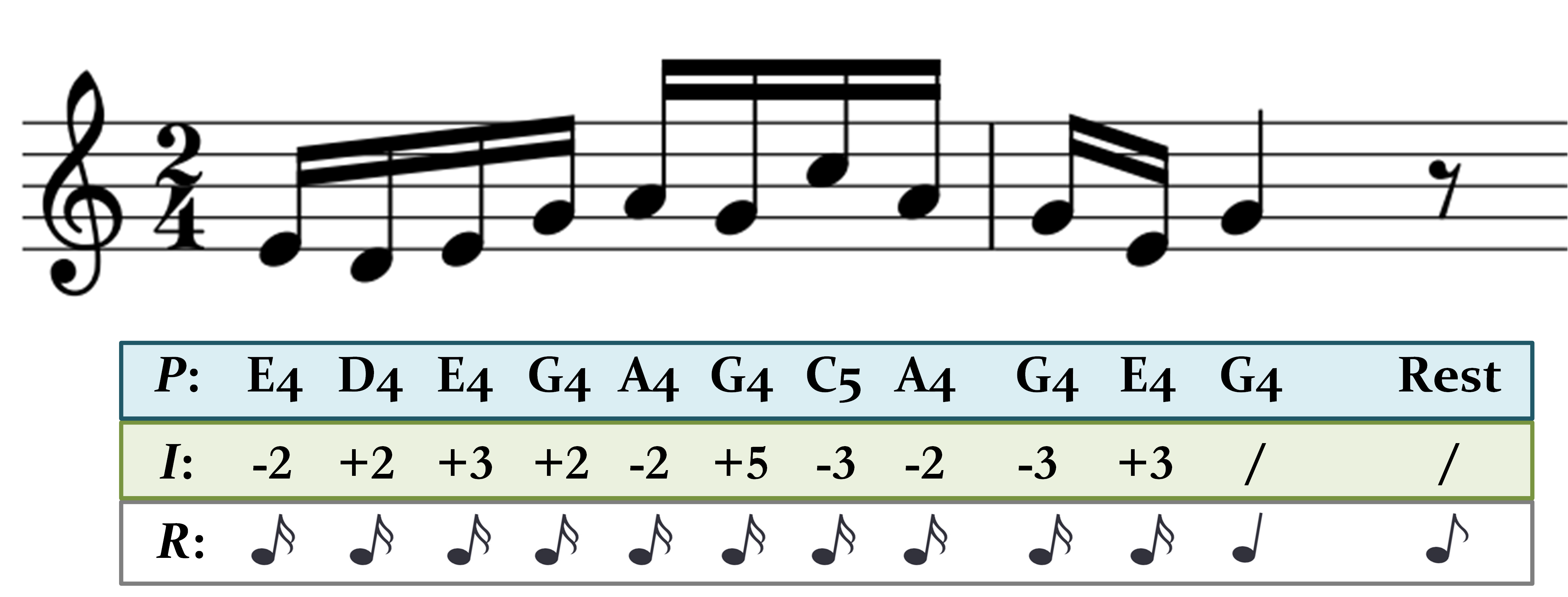}
	\caption{Chinese folk songs representation including pitch sequence, interval sequence, rhythm sequence.}
	\label{fig:1}
\end{figure}

The monophonic folk songs $M$ can be represented as a sequence of note tokens, which is a combination of its pitch, interval and rhythm. Pitch and rhythm are essential information for music. The interval is an important indicator to distinguish the regional music feature, especially for Han Chinese folk songs\cite{Han1989}. The detail processing is described as below and shown in Fig.~\ref{fig:1}.

\begin{itemize}
\item \textbf{Pitch Sequence} $P$: Sequence of pitch tokens which consists of the pitch type presented in melody sequence. Rest note is assigned a special token.
	
\item \textbf{Interval Sequence} $I$: Sequence of interval tokens derived from $P$. Each interval token is represented as a deviation between the next pitch and current pitch in step of semitone.
	
\item \textbf{Rhythm Sequence} $R$: Sequence of duration tokens comprised of the duration type presented in melody sequence.
\end{itemize}

\subsection{Model}

As mentioned in Section 1, the regional characteristics of Chinese folk songs are mainly reflected in their pitch patterns and rhythm patterns. In some areas, the regional characteristics of folk songs are more dependent on pitch feature, while the rhythm patterns in some areas are more distinctive. For example, in terms of pitch, folk songs in northern Shaanxi tend to use perfect forth, the Hunan folk songs often use the combination of major third and minor third \cite{Miao1985}, while Uighur folk songs employ the non-pentatonic scale. In terms of rhythm, Korean folk songs have their special rhythm system named \emph{Jangdan}, while Mongolian \emph{Long Song}s generally prefer long duration notes \cite{Du2014}. 

Inspired by the above observations, it is necessary to further refine the style of folk songs both in pitch and rhythm. Therefore, we propose a VAE-based model to separate pitch space and rhythm space, and further disentangle the music style and content space from pitch and rhythm space, respectively. 

\subsubsection{VAE and its Latent Space Division}

The VAE introduces a continuous latent variable $z$ from a Gaussian prior $p_{\theta}(z)$, and then generates sequence $x$ from the distribution $p_{\theta}(x|z)$ \cite{KingmaW13}. Concisely, a VAE includes an encoder $q_{\phi}(z|x)$, a decoder $p_{\theta}(x|z)$  and latent variable $z$. The loss function of VAE is 

\begin{equation}
J(\phi, \theta) = -\mathbb{E}_{q_{\phi}(z|x)}[{\rm log}p_{\theta}(x|z)] + \beta KL(q_{\phi}(z|x)\|p_{\theta}(z))
\label{eq:1}
\end{equation}

where the first term denotes reconstruction loss, and the second term refers to the Kullback-Leibler (KL) divergence, which is added to regularize the latent space. Weight $\beta$  is a hyperparameter to balance the two loss terms. By setting $\beta<1$, we can improve the generation quality of the model \cite{Higgins2017}. $p_{\theta}(z)$  is the prior and generally obeys the standard normal distribution, i.e., $p_{\theta}(z) = \mathcal{N}(0,I)$. The posterior approximation $q_{\phi}(z|x)$ is parameterized by encoder which is also assumed to be Gaussian and reparameterization trick is used to acquire its mean and variance.

\begin{figure}[htb]
	\centering
	\includegraphics[width=4.4in]{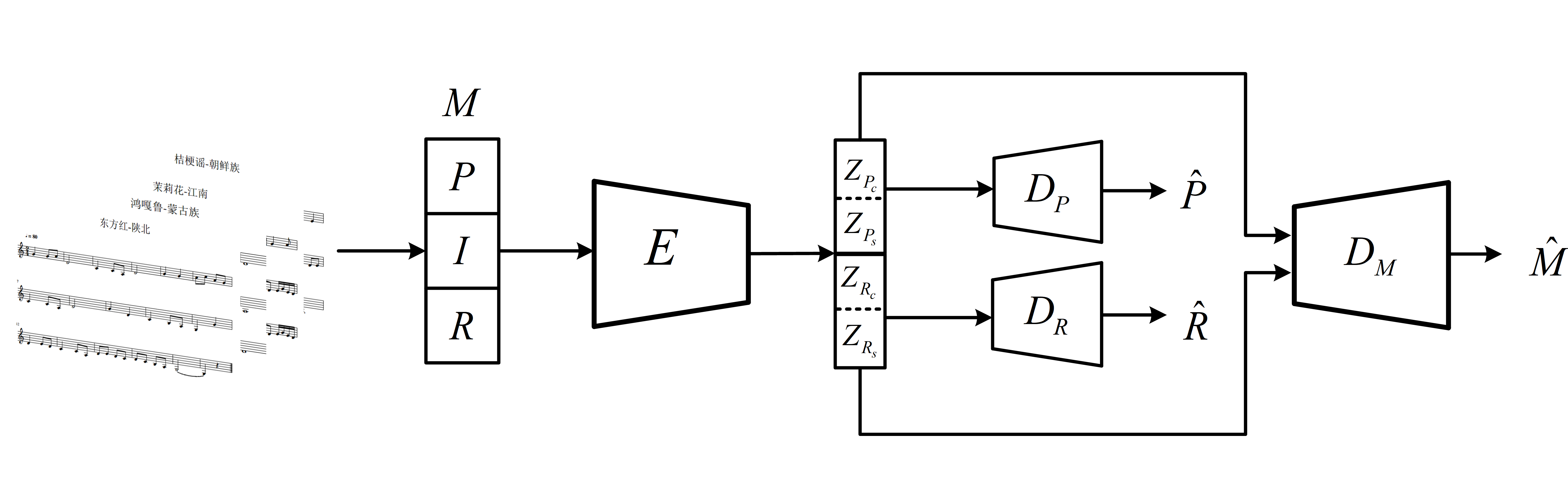}
	\caption{Architecture of our model, it consists of melody encoder $E$, pitch decoder $D_P$, rhythm decoder $D_R$ and melody decoder $D_M$.}
	\label{fig:2}
\end{figure}

With the labeled data, we can disentangle the latent space of VAE in a way that different parts of the latent space correspond to different external attributes, which can enable the generation process in a more controllable way. In our case, we assume that the latent space can be firstly divided into two independent parts, i.e., \emph{pitch variable} and \emph{rhythm variable}. The pitch variable learns the pitch features of Chinese folk songs, while rhythm variable captures the rhythm patterns. Further, we assume both the pitch variable and rhythm variable consist of two independent parts, which refer to music \emph{style variable} and music \emph{content variable}, respectively. 

Specifically, given a melody sequence $M=\{m_1,m_2,\cdots,m_n\}$ as the input sequence with $n$ tokens (notes), where $m_k$ denotes the feature combination of the corresponding pitch token $p_k$, interval sequence $i_k$ and rhythm sequence $r_k$, we firstly encode $M$ and obtain four latent variables from the linear transformation of the encoder’s output. The four latent variables are pitch style variable $Z_{P_s}$, pitch content variable $Z_{P_c}$, rhythm style variable $Z_{R_s}$  and rhythm content variable $Z_{R_c}$, respectively. Then, we concatenate $Z_{P_s}$ and $Z_{P_c}$ into the total pitch variable $Z_{P}$, which is used to predict the pitch sequence $\hat{P}$. The same operation is launched in rhythm variable to predict $\hat{R}$. Finally, all latent variables are concatenated to predict the total melody sequence  $\hat{M}$. The architecture of our model is shown in Fig.~\ref{fig:2}.

Based the above assumption and operation, it is easy to extend the basic loss function:
\begin{equation}
J_{vae} = H(\hat{P},P) + H(\hat{R},R) + BCE(\hat{M},M) + \beta KL_{total}
\label{eq:2}
\end{equation}

where $H(\cdot,\cdot)$ and $BCE(\cdot,\cdot)$ denote the cross entropy and binary cross entropy between prediction values and target values, respectively, and $KL_{total}$  denotes the sum KL loss of the four latent variables.

\subsubsection{Adversarial Training for Latent Spaces Disentanglement}

\begin{figure}[htb]
	\centering
	\includegraphics[width=4.5in]{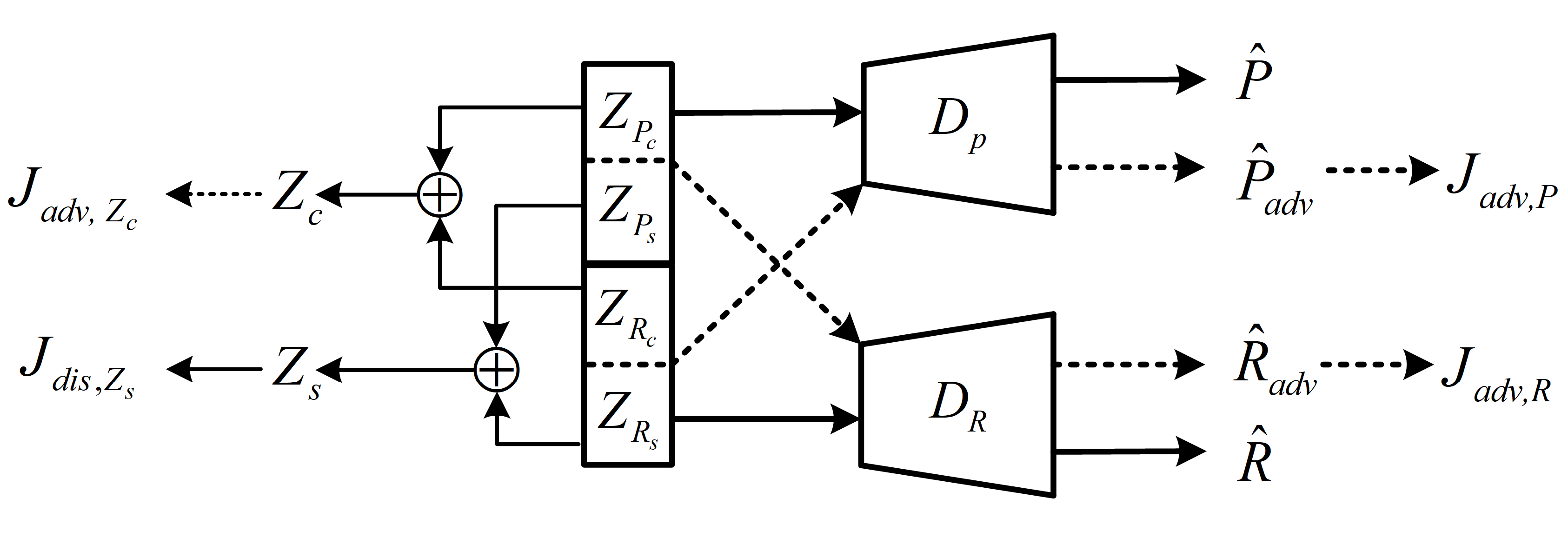}
	\caption{Detail processing of latent spaces disentanglement. The
		dashed lines indicate the adversarial training parts.}
	\label{fig:3}
\end{figure}

Here, we propose an adversarial training based method to conduct the disentanglement of pitch and rhythm, music style and content. The detail processing is shown in Fig.~\ref{fig:3}.

As shown in Fig.~\ref{fig:2}, we use two parallel decoders to reconstruct pitch sequence and rhythm sequence, respectively. Ideally, we expect the pitch variable $Z_P$ and rhythm variable $Z_R$ should be independent of each other. However, the pitch feature may be implicit in rhythm variables actually, vice versa, since the two variables are sampled from the same encoder output. 

In order to separate the pitch and rhythm variable explicitly, the temporal supervision is employed in the separation of pitch and rhythm, which is similar to the work of disentangled representation for pitch and timbre \cite{IJCAI2019}. Specifically, we feed the latent variable to the wrong decoder deliberately and force the decoder to predict nothing, i.e., all zero sequence, resulting in the following two loss terms based on cross entropy:

\begin{equation}
J_{adv,P} = -\Sigma[\mathbf{0}\cdot {\rm log}\hat{P}_{adv} + (1-\mathbf{0})\cdot {\rm log}(1-\hat{P}_{adv})]
\label{eq:3}
\end{equation}

\begin{equation}
J_{adv,R} = -\Sigma[\mathbf{0}\cdot {\rm log}\hat{R}_{adv} + (1-\mathbf{0})\cdot {\rm log}(1-\hat{R}_{adv})]
\label{eq:4}
\end{equation}

where $\mathbf{0}$ denotes all zero sequence, $`\cdot$' denotes the element-wise product.

For the disentanglement of music style and content, we firstly obtain the total music style variable $Z_s$ and content variable $Z_c$:

\begin{equation}
Z_s = Z_{P_s} \oplus Z_{R_s}, Z_c = Z_{P_c} \oplus Z_{R_c}
\label{eq:5}
\end{equation}

where $\oplus$ means the concatenate operation. 

Then two classifiers are defined to force the separation of style and content in the latent space using the regional information. The style classifier ensures the style variable is discriminative for regional label, while the adversary classifier force the content variable is not distinctive for regional label. For style classifier is trained with the cross entropy defined by
\begin{equation}
J_{dis, Z_s} = -\Sigma y {\rm log}p(y|Z_s)
\label{eq:6}
\end{equation}

where $y$ denotes the ground truth, $p(y|Z_s)$ is the predicted probability distributions from style classifier.

For adversary classifier, we train it by maximizing the empirical entropy of the adversary classifier's prediction \cite{FuTPZY18}\cite{Vineet2018}. The training processing is divided into two steps. Firstly, the parameters of the adversary classifier are trained independently, i.e., the gradients of the classifier don't propagate back to VAE. Secondly, we compute the empirical entropy based on the output from adversary classifier as defined by

\begin{equation}
J_{adv, Z_c} = -\Sigma p(y|Z_c) {\rm log}p(y|Z_c)
\label{eq:7}
\end{equation}
where $p(y|Z_c)$ is the predicted probability distributions from adversary classifier. 

In summary, the overall training objective of our model is the minimization the loss function defined by

\begin{equation}
J_{total} = J_{vae} + J_{adv,P} + J_{adv, R} + J_{dis, Z_s} - J_{adv, Z_c}
\label{eq:8}
\end{equation}

\section{Experimental Results and Analysis}
\subsection{Datasets and Preprocessing}

The lack of large-scale Chinese folk song datasets makes it impossible to apply deep learning methods for automatic generation and analysis of Chinese music. Therefore, we digitize more than 2000 Chinese folk songs in MIDI format from the record of \emph{Chinese Folk Music Integration}\footnote{\emph{Chinese Folk Music Integration} is one of the major national cultural project leaded by the former Ministry of Culture, National Ethnic Affairs Commission and Chinese Musicians Association from 1984 to 2001. This set of book contains more than 40000 selected folk songs of different nationalities. The project website is \url{http://www.cefla.org/project/book }.}. These songs contain Han folk songs from Wu dialect district, Xiang dialect district\footnote{According to the analysis of Han Chinese folk songs \cite{Han1989}\cite{Du1993}, the folk song style of each region is closely related to the local dialects. Therefore, we classify Han folk songs based on dialect divisions. Wu dialect district here mainly includes Southern Jiangsu, Northern Zhejiang and Shanghai. Xiang dialect district here mainly includes Yiyang, Changsha, Hengyang, Loudi and Shaoyang in Hunan province.}  and northern Shaanxi, as well as three ethnic minority folk songs of Uygur in Xinjiang, Mongolian in Inner Mongolia and Korean in northeast China.

All melodies in datasets are transposed to C key. We use the Pretty-midi python toolkit \cite{raffel2014intuitive} to process each MIDI file, and count the numbers of pitch token, interval token and rhythm token as the feature dimension of the corresponding sequence, which are 40, 46 and 58, respectively. Then pitch sequence, interval sequence and rhythm sequence are extracted from raw notes sequence with the overlapping window of length 32 tokens and a hop-size of 1. Finally, we get 65508 ternary sequences in total. The regional labels of the token sequences drawn from the same song are consistent.

\subsection{Experimental Setup}

\begin{figure}[htb]
	\centering
	\includegraphics[width=2.8in]{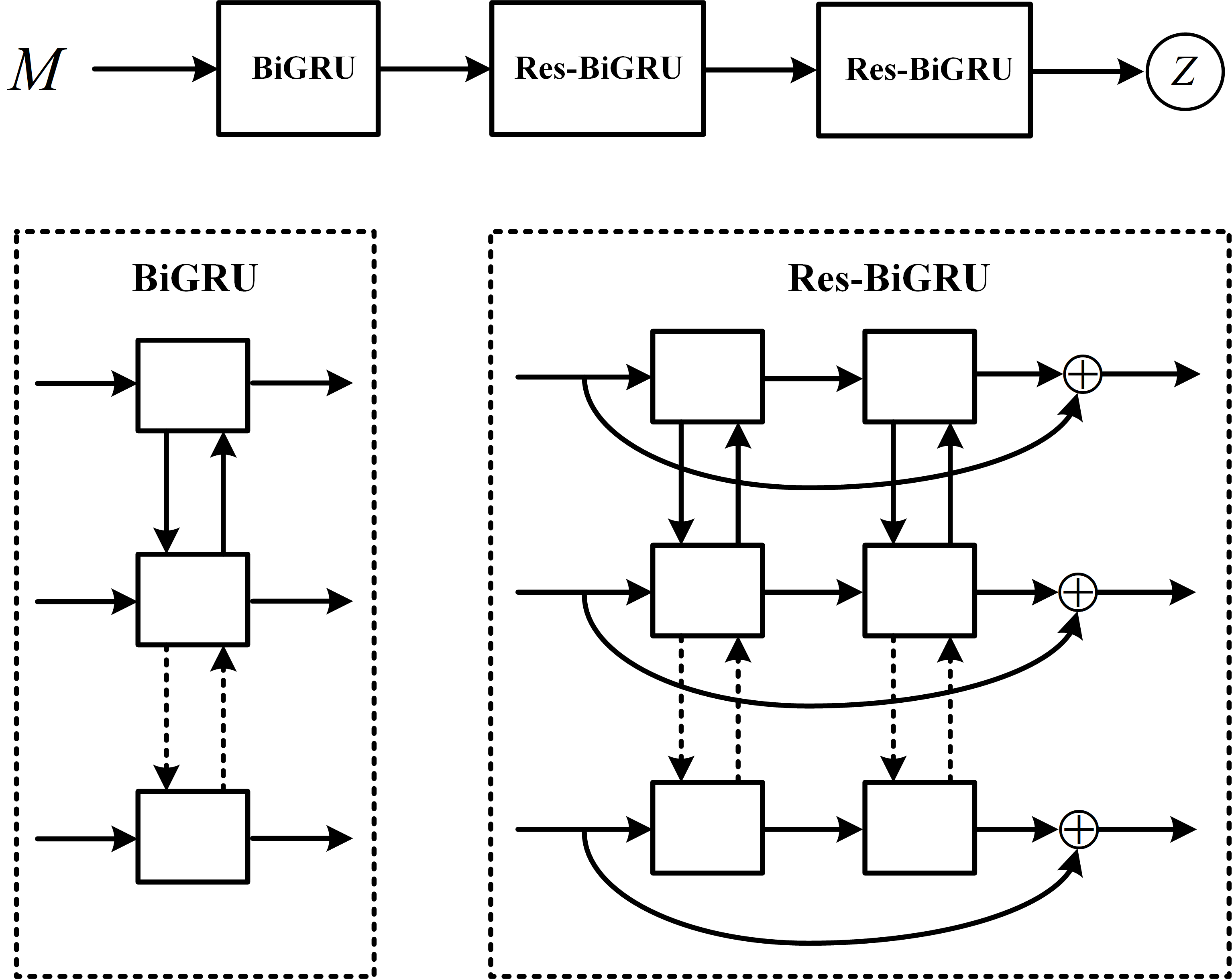}
	\caption{Encoder with residual connections.}
	\label{fig:4}
\end{figure}

In order to extract melody feature into latent space effectively, we employ a bidirectional GRU model with the residual connection \cite{HeZRS16} as encoder, which is illustrated in Fig.~\ref{fig:4}. The decoder is a normal two-layers GRU. All recurrent hidden size in this paper is 128. Both style classifier and adversary classifier are one-layer linear layer with Softmax function. The size of pitch style variable and rhythm style variable is set to 32, while the size of pitch content variable and rhythm content variable is 96. During training period, the KL term coefficient $\beta$ increases from 0.0 to 0.15 linearly to alleviate the impact of posterior collapse.

Adam optimizer is employed with the initial learning rate of 0.01 for VAE training, and vanilla SGD optimizer with the initial learning rate of 0.005 for classifiers. All test models are trained for 30 epochs and the size of mini-batch is set to 50.

\subsection{Evaluation and Results Analysis}

To evaluate the generated music, we employ the following metrics from objective and subjective perspectives.

\begin{itemize}
	\item \textbf{Reconstruction Accuracy}: We calculate the accuracy between the target notes sequence and reconstructed notes sequence on our test set to evaluate the music generation quality.
	\item \textbf{Style Recognition Accuracy}: We train a separate style evaluation classifier using the architecture in Fig.~\ref{fig:4} to predict the regional style of the tunes that are generated using different latent variables. The classifier achieves a reasonable regional accuracy on the independent test set, which is up to 82.71\%. 
	\item \textbf{Human Evaluation}:  As human should be the ultimate judge of creations, human evaluations are conducted to overcome the incoordinations between objective metrics and user studies. We invite three experts who are well educated and expertise in Chinese music. Each expert is asked to listen to the random selected five folk songs of each region on-site, and rate each song on a 5-point scale from 1 (very low) to 5 (very high) according to the following two criteria: a) \emph{Musicality}: Does the song have a clear music pattern or structure? b) \emph{Style Significance}: Does the songs' style match the given regional label?
\end{itemize}

\begin{table}[htb]
	\centering
	\caption{Results of automatic evaluations.}
	\begin{tabular}{ccc}
		\hline
		Objectives & Reconstruction Accuracy  & Style Recognition Accuracy  \\
		\hline
		$J_{vae}$ &  0.7684 & 0.1726/0.1772/0.1814 \\
		\hline
		$J_{vae}, J_{adv,P,R}$ & 0.7926  & 0.1835/0.1867/0.1901 \\
		\hline
		$J_{vae}, J_{adv, P,R}, J_{adv, Z_c}$ &  0.7746  & 0.4797/0.4315/0.5107 \\
		\hline
		$J_{vae}, J_{adv, P,R}, J_{dis, Z_s}$ & \textbf{0.8079}   & 0.5774/0.5483/0.6025 \\
		\hline
		$J_{total}$ &  0.7937   &\textbf{0.6271}/\textbf{0.5648}/\textbf{0.6410} \\
		\hline
	\end{tabular}
	\label{tab:1}
\end{table}

Tab~\ref{tab:1} shows all evaluation results of our models. The three values in the third column denote the accuracies derived from the following three kinds of latent variables: a) the concatenation of pitch style variable $Z_{P_s}$ and a random variable sampled from standard normal distribution; b)  the concatenation of rhythm style variable $Z_{R_s}$ and the random variable; c) the concatenation of total style variable $Z_s$ and the random variable. $J_{adv,P,R}$ denotes the sum of $J_{adv,P}$ and $J_{adv, R}$.

The model with $J_{total}$ achieves the best results in style recognition accuracy and a sub-optimal result in reconstruction accuracy. The model without any constraints performs poorly on the two objective metrics. The addition of $J_{adv,P,R}$ improves the the reconstruction accuracy but fails to bring meaningful improvement to style classification. With the addition of either $J_{adv, Z_c}$ or $J_{dis, Z_s}$, all the three recognition accuracies improve a lot, which indicates that the latent spaces are disentangled into style and content subspaces as expected. Moreover, only employing pitch style or rhythm style for style recognition can also obtain fair results, demonstrating the disentanglement of pitch and rhythm is effective.

\begin{figure}[htb]
	\centering
	\includegraphics[width=4.6in]{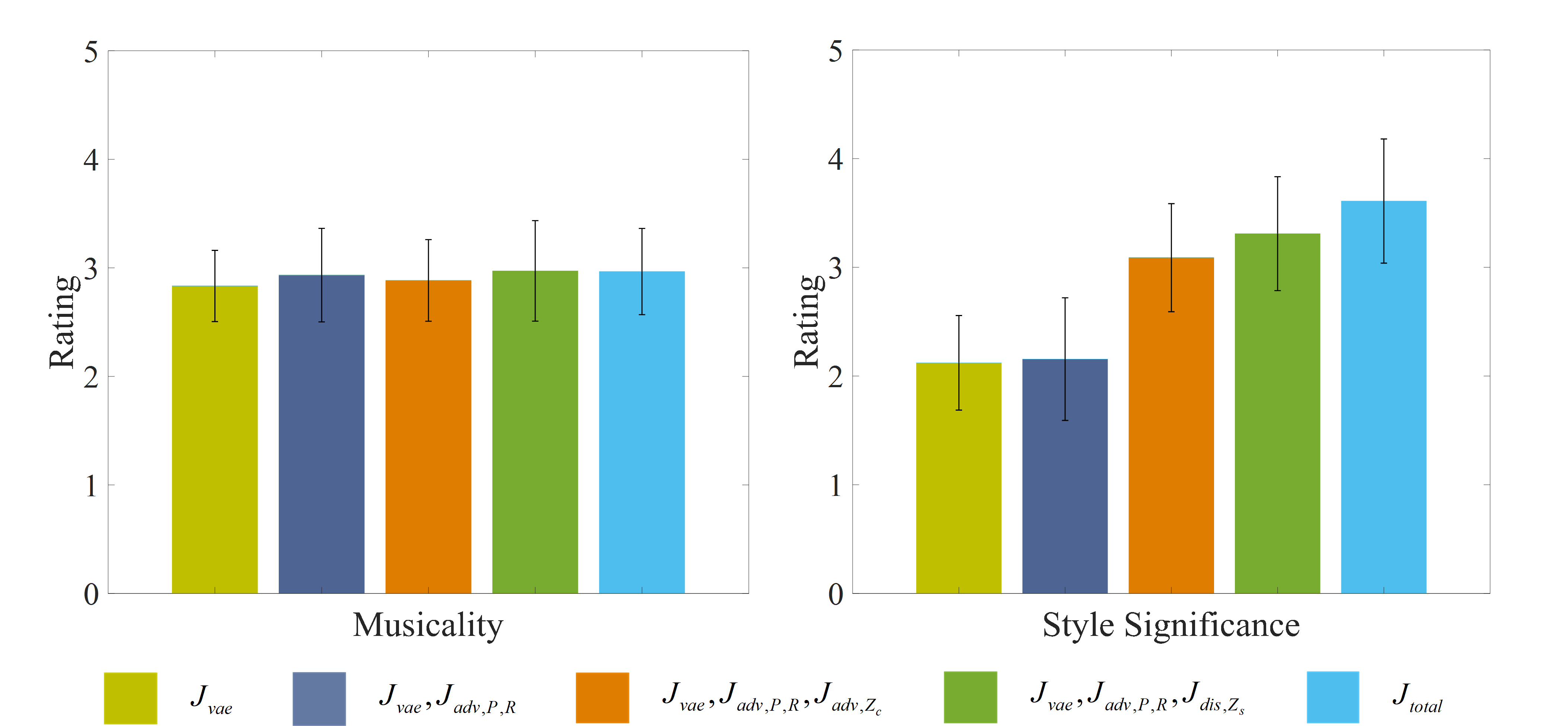}
	\caption{Results of human evaluations including musicality and style significance. The heights of bars represent means of the ratings and the error bars represent the standard deviation.}
	\label{fig:5_add}
\end{figure}

The result of human evaluations is shown in Fig.~\ref{fig:5_add}. In terms of musicality, all test models have similar performance, which demonstrates the addition of extra loss function has no negative impact on the generation quality of original VAE. Moreover, the model with total objectives $J_{total}$ performs significantly better than other models in terms of style significance (two-tailed $t$-test, $p<0.05$), which is consistent with the results in Tab~\ref{tab:1}.

\begin{figure*}[!t]
	\centering
	\subfigure[Pitch style latent space ]{\includegraphics[width=2.2in]{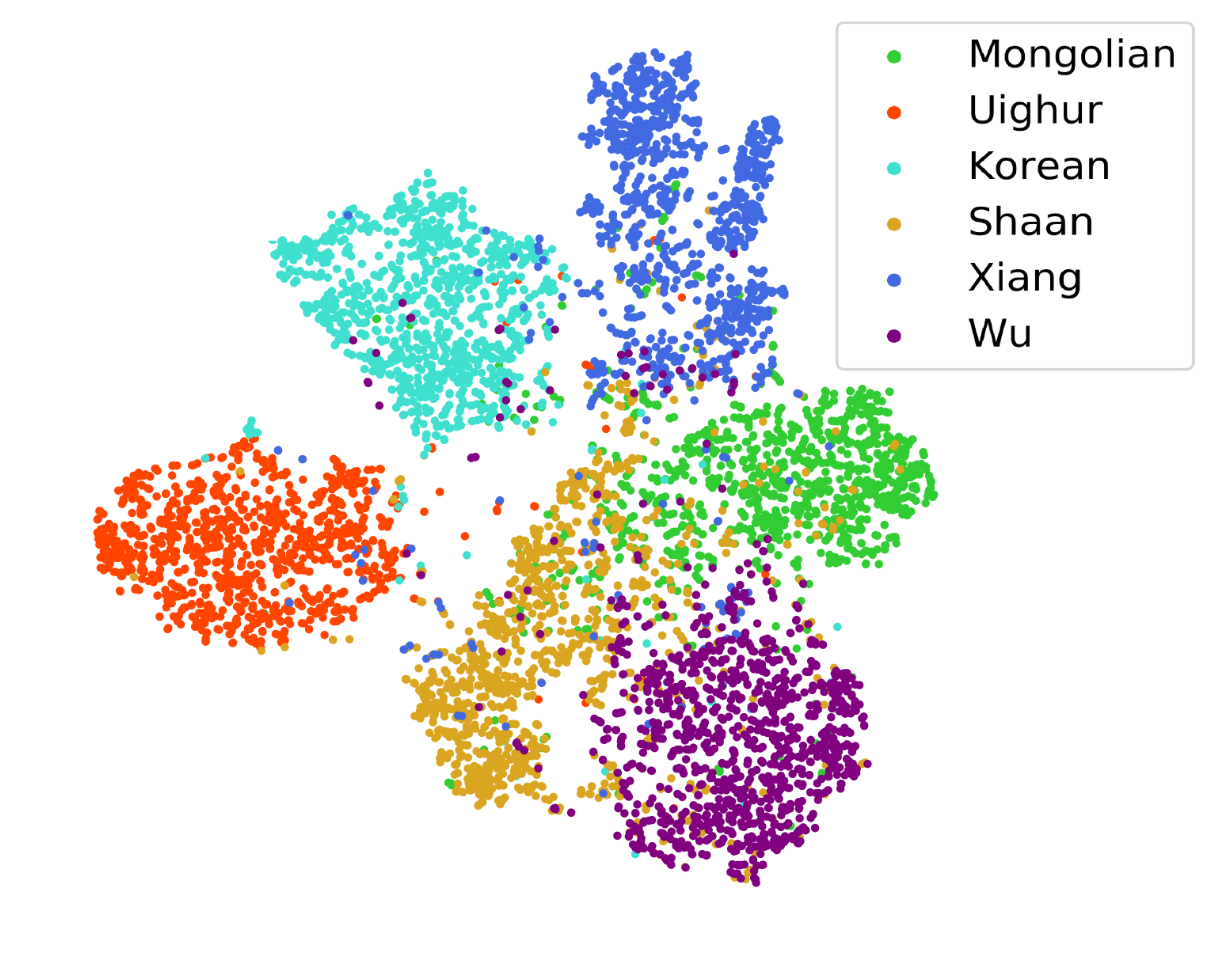} \label{fig:5a}}~~
	\subfigure[Rhythm style latent space ]{\includegraphics[width=2.2in]{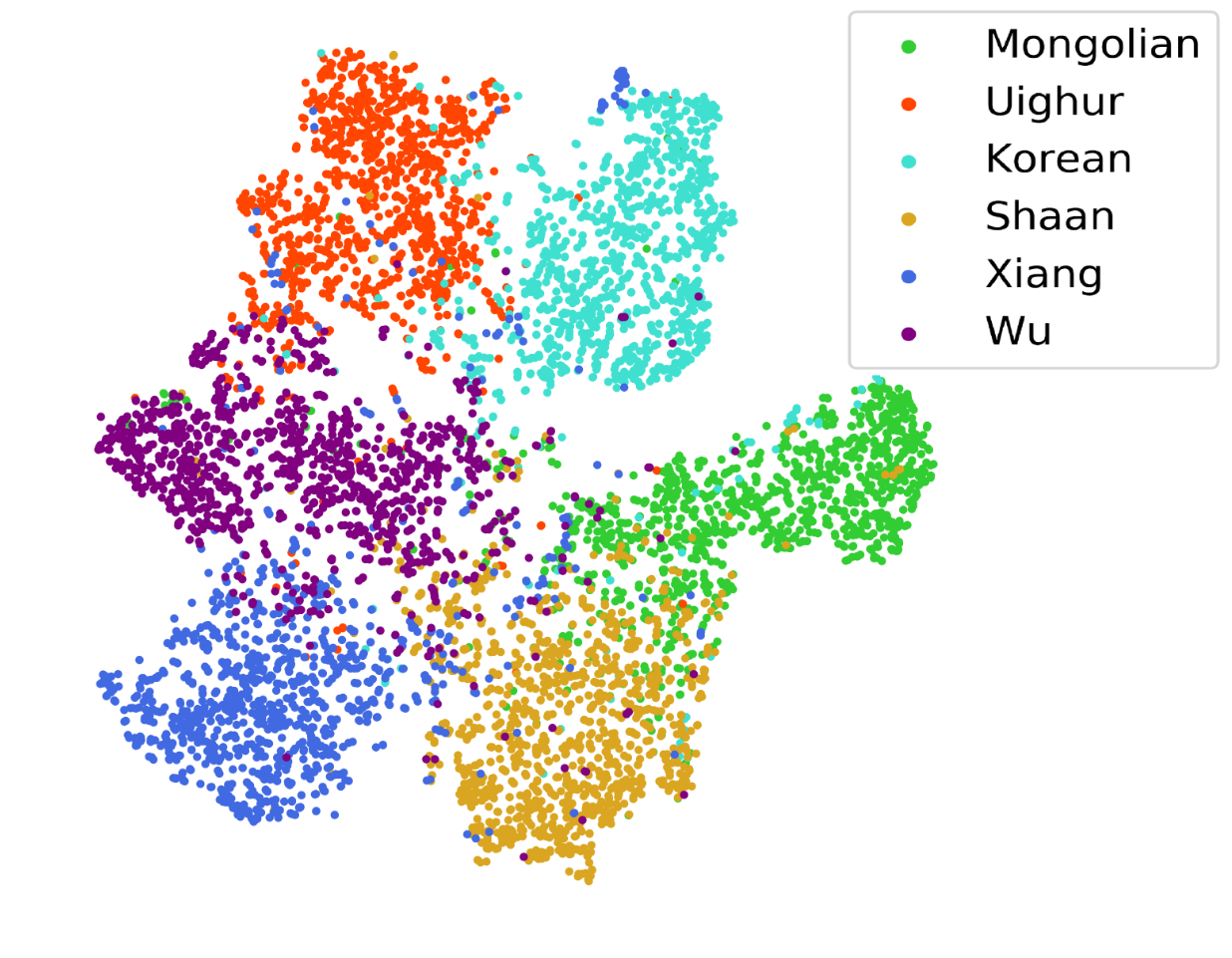} \label{fig:5b}}
	\subfigure[Total style latent space]{\includegraphics[width=2.2in]{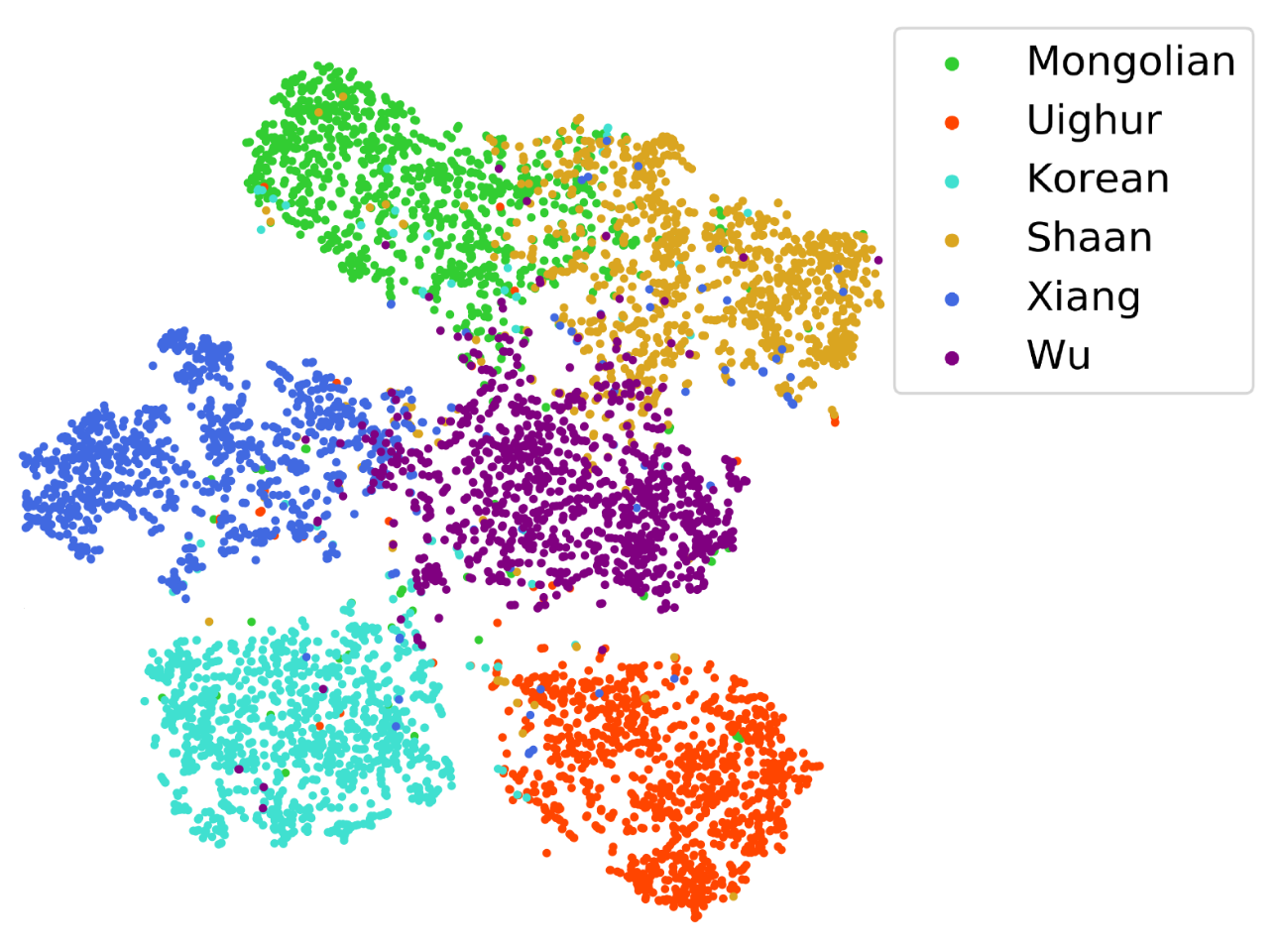} \label{fig:5c}}~~
	\subfigure[Total content latent space ]{\includegraphics[width=2.2in]{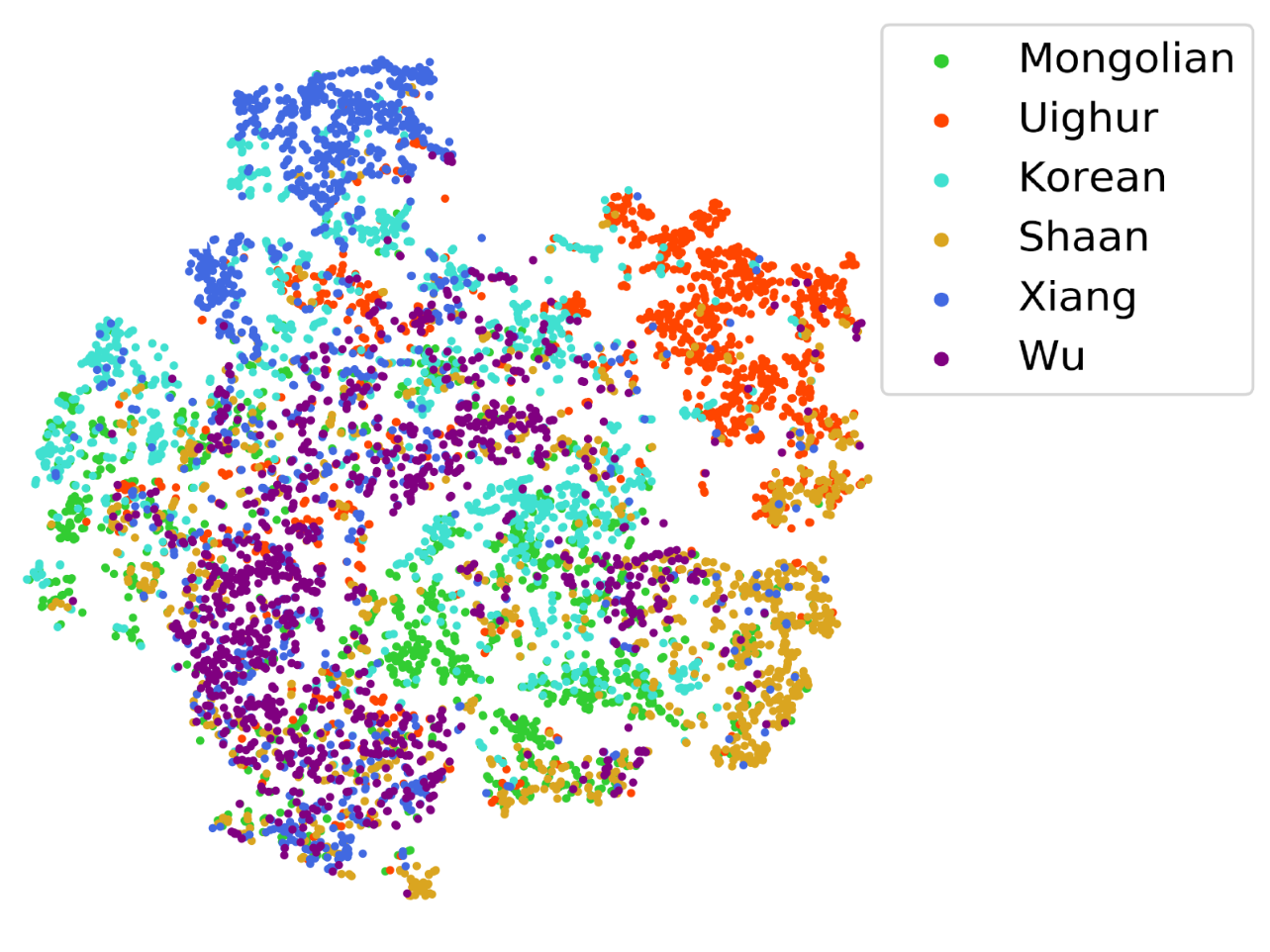} \label{fig:5d}}
	\caption{t-SNE visualization of model with $J_{total}$.}
	\label{fig:5}       
\end{figure*}

Fig.~\ref{fig:5} shows the t-SNE visualization\cite{maaten2008visualizing} of our model with $J_{total}$. We can observe that music with different regional labels is noticeably separated in the pitch style space, rhythm style space and total style space, but looks chaos in content space.This further demonstrates the validity of our proposed methods to disentangle the pitch, rhythm, style and content.

\begin{figure}[htb]
	\centering
	\includegraphics[width=4.8in]{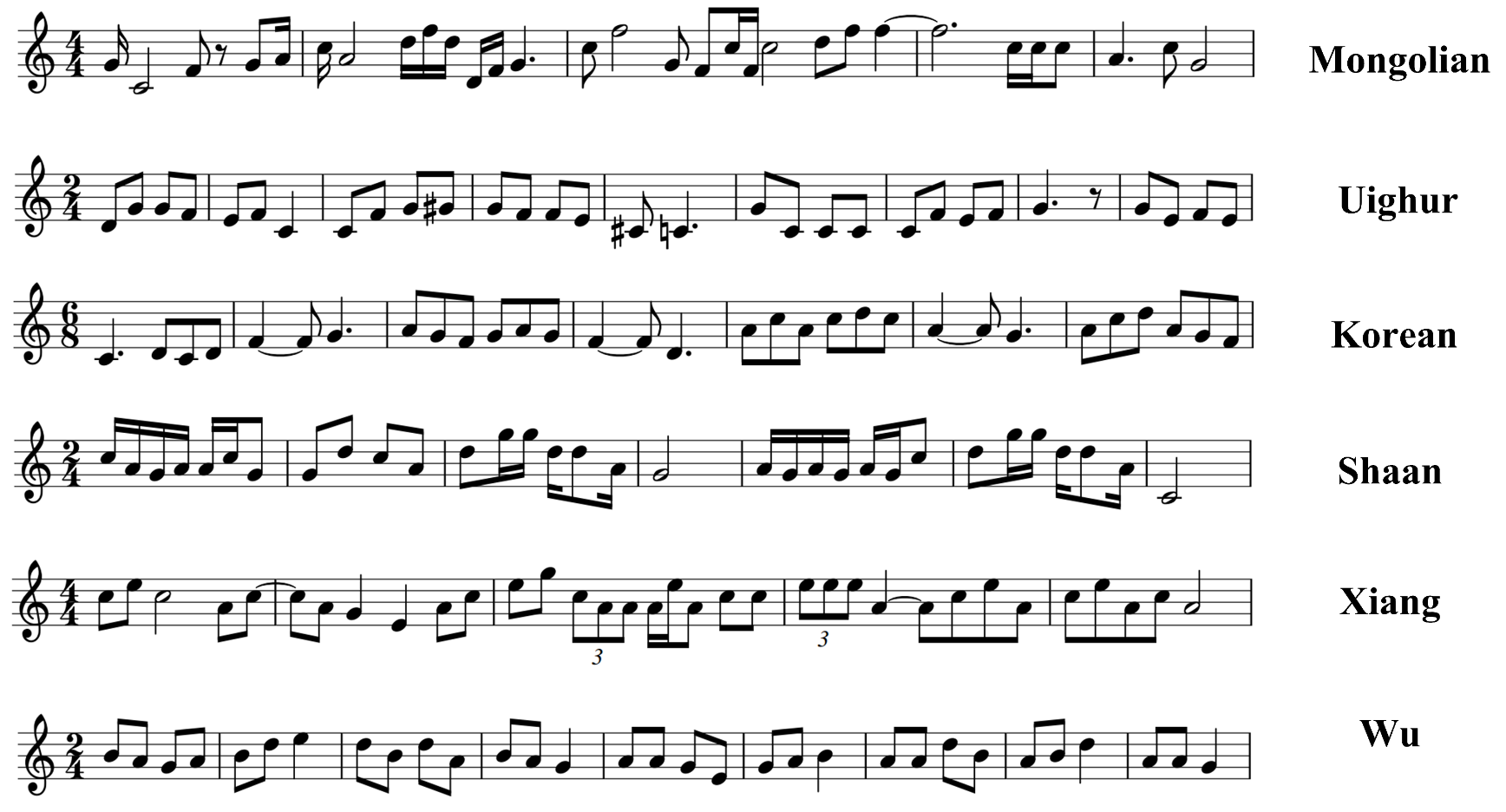}
	\caption{Examples of folk songs generation given regional labels. In order to align each row, the scores of several regions are not completely displayed.}
	\label{fig:6}
\end{figure}

Finally, we present several examples \footnote{Online Supplementary Material: \url{https://csmt201986.github.io/mgvaeresults/}.} of generating folk song with given regional labels with our methods in Fig.~\ref{fig:6}. As seen, we can create novel folk songs with dominated regional features such as long duration notes and large interval in Mongolian songs, the combination of major third and minor third in Hunan folk songs, and so on. However, there are still several failed examples. For instance, few generated songs repeat same melody pattern. More commonly, some songs don't show the correct regional feature, especially when the given regions belong to Han nationality areas. This may due to the fact that folk tunes in those regions share the same tonal system.

\section{Conclusion}

In this paper, we focus on how to capture the regional style of Chinese folk songs and generate novel folk songs with specific regional labels. We firstly collect a database including more than 2000 Chinese folk songs for analysis and generation. Then, inspired by the observation of the regional characteristics in Chinese folk songs, a model named MG-VAE based on adversarial learning is proposed to disentangle the pitch variable, rhythm variable, style variable and content variable in the latent space of VAE. Three metrics containing automatic and subjective evaluation in our experiments are used to evaluate the proposed model. Finally, the experimental results and t-SNE visualization show that the disentanglement of the four variables is successful and our model is able to generate folk songs with controllable regional style. In the future, we plan to expand the proposed model to generate longer melody sequence using more powerful model like Transformers, and explore the evolution of tune families like \emph{Mo Li Hua Diao}, \emph{Chun Diao} among different regions.

\bibliographystyle{spmpsci}     
\bibliography{mgvae}

\end{document}